\documentclass[12pt]{article}
\usepackage[dvips]{graphicx}
\begin{document}
\title{Liquid Argon Ionization Detector for Double Beta Decay Studies}
\author{V.D. Ashitkov$^1$, A.S. Barabash$^1$, S.G. Belogurov$^1$,
G. Carugno$^2$,
\\ S.I. Konovalov$^1$, F. Massera$^3$, G.
Puglierin$^2$, R.R. Saakyan$^{1,4}$,\\
V.N. Stekhanov$^1$, V.I. Umatov$^1$,
\\[0.4cm]
$^1${\small Institute of Theoretical and Experimental Physics, B.\
Cheremushkinskaya 25,}\\
{\small 117259 Moscow, Russia} \\
$^2${\small Dipartimento di Fisica e INFN, Universita di Padova,
via Marzolo 8,}\\
{\small I-35131 Padova, Italy} \\
$^3${\small INFN, Sezione di Bologna, 40126, via Berti Pichat, 6/2
(Bologna), Italy}\\
$^4${\small Department of Physics and Astronomy,
University College London,}\\
{\small Gower Street, LONDON, WC1E 6BT UK }\\
}
\date{ }
\maketitle

\begin{abstract}
A multisection liquid argon ionization detector was developed by
the DBA collaboration to study the double beta- decay of
$^{100}$Mo. The experiment was carried out in the Gran Sasso
underground laboratory in Italy. The detector design and main
characteristics are described. The $\beta \beta (2\nu)$ decay of
$^{100}$Mo was observed and its half-life measured: $T_{1/2}=[7.2
\pm 0.9(stat) \pm 1.8(syst)] \times 10^{18}$ yr. Limits on the
0$\nu $ and 0$\nu \chi ^{0}$ modes of the decay were obtained:
$T_{1/2}> 8.4(4.9) \times 10^{21}$ yr and $T_{1/2}> 4.1(3.2)\times
10^{20}$ yr at 68 \% (90 \%) C.L., respectively. In addition the
upper limits on the $^{42}$Ar content and $^{222}$Rn activity in
liquid Ar were found to be $4.3 \times 10^{-21}$ g/g and $1.2
\times 10^{-3}$ Bq/kg, respectively.

\end{abstract}

\section{Introduction}
The current interest in $\beta \beta (0\nu )$ decay is due to the
fact that its existence is closely linked to the following
fundamental aspects of particle physics
\cite{KLAP98,FEAS01,VER02}:
 \\-lepton number non-conservation;
 \\-existence of neutrino mass and its nature;
 \\-existence of right-handed currents in the electroweak
 interactions;
 \\-existence of the Majoron;
 \\-the Higgs sector structure;
 \\-existence of leptoquarks;
 \\-heavy sterile neutrino and composite neutrino.
 \\All these questions lie beyond the scope of the Standard Model of
electroweak interactions and so the detection of $\beta \beta
(0\nu)$ decay will give rise to {"}new physics{"}. The primary
interest in this process is the question of neutrino mass: \ -the
detection of $\beta \beta (0\nu )$ decay will allow to establish
the absolute neutrino mass scale and will indicate that this mass
is of Majorana type. The recent results from atmospheric
\cite{KAJ01}, solar \cite{CLEV98,ABD99,HAM99,AHM01,FUK01}, and
reactor \cite{EGU03} neutrino experiments indicate that neutrinos
have mass and they mix. This very exciting result has greatly
renewed the interest in $\beta \beta(0\nu )$ decay since the
oscillation experiments show that the next generation of double
beta decay experiments will have a good chance of measuring the
absolute mass of the neutrino. To date, only lower bounds on
half-lives ($T_{1/2}0\nu$) have been obtained experimentally for a
number of double beta decay isotopes. These bounds are used to
deduce upper limits on the Majorana neutrino mass, the
right-handed current admixture parameter, the Majoron-Majorana
neutrino coupling constant etc. However, uncertainties in nuclear
matrix element (NME) calculations do not allow reliable limits to
be placed on these fundamental values. In this context the
detection of two-neutrino double beta decay $\beta \beta (2\nu )$
becomes of particular importance because information on
experimental values of NME (2$\nu $) for different nuclei enables
a more accurate calculation of both NME (2$\nu $) and NME (0$\nu
$).
\\A liquid argon detector \cite{BAR86,ASH01} was designed by
the DBA collaboration to study the double-beta decay of
$^{100}$Mo. The $^{100}$Mo isotope was selected for the study
because of its large $\beta \beta $ transition energy (E$_{\beta
\beta }$ = 3033 keV). In 1996 the experimental setup was installed
in the Gran Sasso underground laboratory (Italy). In this work we
describe the technique used in the experiment. Special attention
is paid to the design of the experimental setup, calibration
procedure and performance of the detector components.
\section{Experimental setup}
\begin{figure}
\begin{center}
\includegraphics[width=10cm]{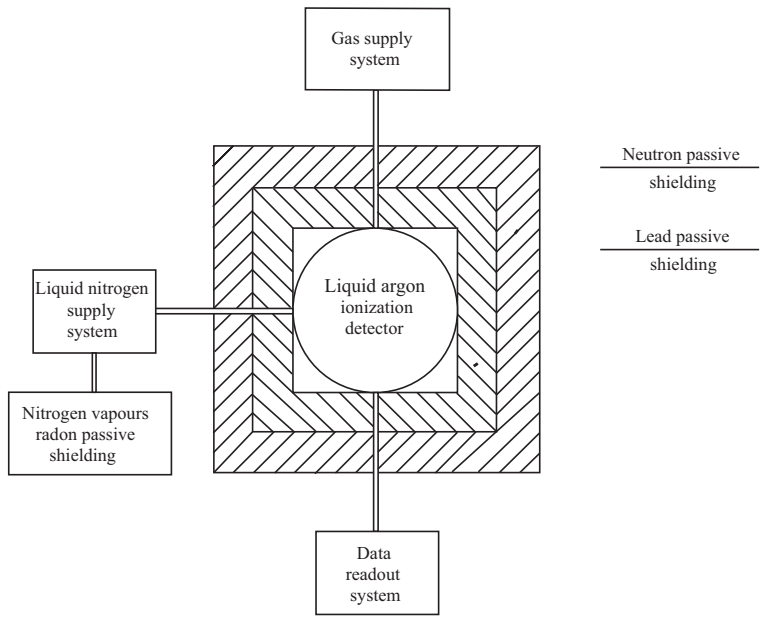}
\caption{Sketch of the experimental setup.}
\label{exsetup}
\end{center}
\end{figure}
The experiment was carried out in the Gran Sasso Underground
Laboratory located at a depth of 3500 m w.e. (meters of water
equivalent). The experimental setup (Fig.~\ref{exsetup})
consists of a liquid argon ionization detector enclosed in passive
lead shielding, a gas supply system, a radon protection system, a
mechanical crane used for detector assembling and disassembling,
electronics and a data acquisition system. The equipment was
mounted on a special concrete platform with dimensions of 4$\times
6\times $0.6 m. To suppress mechanical vibrations the platform was
supported on a rubber sheet. The platform carried a special
{"}house{"} which protected the equipment from the moist
atmosphere underground. The data acquisition and readout system of
the detector was placed in another {"}house{"} located 2 m away
from the platform.

\subsection{Liquid-argon ionization detector}

\begin{figure}
\begin{center}
\includegraphics[width=10cm]{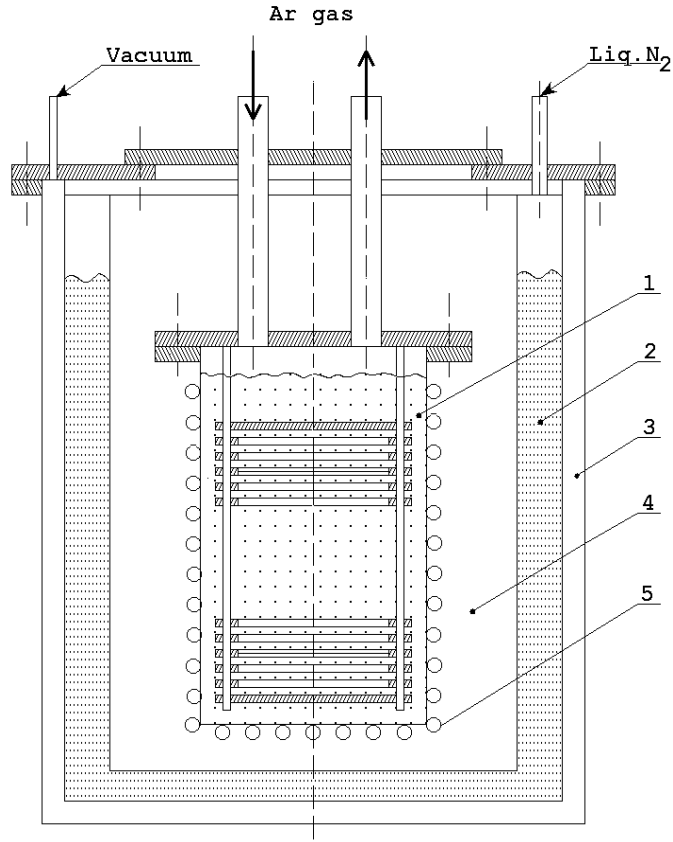}
\caption{Sketch of the liquid-argon ionization detector.}
\label{Chamber}
\end{center}
\end{figure}

The detector is mainly constructed of titanium and its alloys. All
the insulators in the electrode system were made of fluoroplastic.
A sketch of the detector is shown in Fig.~\ref{Chamber}. The
detector was cooled with liquid nitrogen that was fed into a
nitrogen volume (2) from a 4000 l Dewar vessel. The liquid
nitrogen was pumped to the detector through a pipeline, enclosed
in vacuum heat insulation to reduce the liquid nitrogen flow rate,
which was necessary for long period of routine data taking. The
nitrogen volume (2) was thermally insulated from the environment
with a vacuum casing (3). The inner wall of the volume was coated
with 12 layers of 10 $\mu $m-thick aluminized Mylar. This thermal
insulation allowed the liquid nitrogen flow rate to be reduced to
6 l/h. The nitrogen volume (2) was connected to a chamber (4), in
which a vessel (1) with liquid argon was housed, with a tube to
equalize the nitrogen vapor pressure in these volumes. Heaters (5)
located on the outside wall of the vessel (1) regulated the liquid
argon temperature.

\begin{figure}
\begin{center}
\includegraphics[width=8cm]{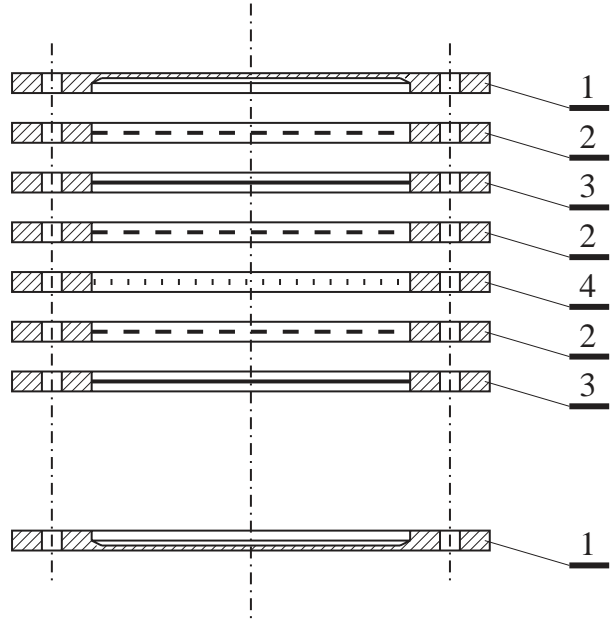}
\caption{Arrangement of the electrodes in the liquid-argon
ionization detector. 1 -  Solid anode; 2 - screening grid; 3 -
molybdenum-foil cathode and 4 -  wired anode.}
\label{Elsy}
\end{center}
\end{figure}

The electrode system that formed the sensitive volume of the
detector was placed in the vessel (1), 40 cm in diameter and 70 cm
high. The active detection portion of the detector was composed of
identical sections (Fig.~\ref{Elsy}) Each section consisted of
two combined flat ionization chambers with screening grids (2) and
a common cathode (3). A foil made of the isotope under study
(molybdenum) was inserted into a circular cathode frame. The
sensitive volume was 30 cm in diameter and 56 cm high. The
detector contained 14 cathodes, 15 anodes, and 28 screening grids.
The grid-anode and grid-cathode gaps were 5.5 and 14.5 mm,
respectively. The anodes (4) had the same design as the grids. The
screening grids and anodes were made with 110 $\mu $m Ni-Cr wires
wound with a 1 mm spacing. The top and the bottom anodes (1) of
the electrode system were solid titanium discs. The high voltage
applied to the cathodes and the grids was -4.8 and -2 kV,
respectively. Charge sensitive preamplifiers converted the charge,
produced on each anode, into a voltage pulse.

\subsection{Isotopes}

The detector cathodes were made of molybdenum foil 50 mg/cm$^{2}$
thick. The four cathodes of enriched molybdenum (98.4 \% of
$^{100}$Mo) were first investigated in three experimental runs of
202, 238, and 313 h duration. To suppress the radioactive
background these cathodes were mounted at the center of the
electrode system and interleaved with cathodes of natural
molybdenum (containing 9.6 \% of $^{100}$Mo). The activity of
radioactive impurities in the molybdenum foil samples was less
then 0.015 Bq/kg for $^{214}$Bi, 0.0015 Bq/kg for $^{208}$Ti and
0.04 Bq/kg for $^{234m}$Pa. The total mass of $^{100}$Mo under
study was 138.7 g. Later, the $^{100}$Mo mass in the detector was
increased to 306 g (eight cathodes of $^{100}$Mo). The interleaved
arrangement of the natural and enriched molybdenum cathodes
allowed a differential measurement to be made. In this method one
can subtract the natural sections spectrum from the enriched
sections spectrum yielding a pure 2$\nu $ effect. The total
measurement time with eight $^{100}$Mo cathodes was 2706 h.

\subsection{Gas supply system}
\begin{figure}
\begin{center}
\includegraphics[width=10cm]{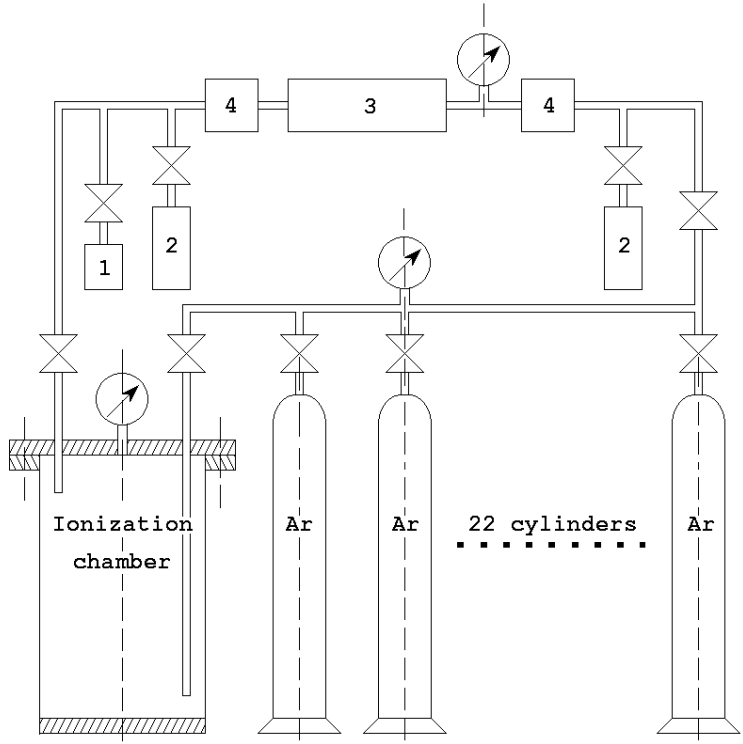}
\caption{Schematic diagram of the gas-supply system. 1- Gas-purity
control system; 2- gas bottles (2 l in volume); 3- purification
system; 4- mechanical filters.}
\label{Gas}
\end{center}
\end{figure}

The gas supply system (Fig.~\ref{Gas}) consisted of 22 40 l
stainless steel bottles with gaseous argon under a pressure of
about 150 bar, a gas purification system (3), and a gas purity
control system (1). Six of these bottles were housed in stainless
steel Dewar vessels. After the measurements they were used as a
cryogenic pump to transfer the liquid argon from the chamber to
the bottles. The gas purification system (3) had a titanium sponge
as its main element due to a low concentration of radioactive
impurities in this getter. For example, a purification system
based on an Oxysorb absorber or a molecular sieve releases
100--1000 times more $^{222}$Rn because the $^{238}$U content in
these materials (approximately 10$^{-6}$ to 10$^{-7}$ g/g) exceeds
that of titanium by the same factor. After a single pass of
gaseous argon through the titanium sponge at a temperature of 850
$^{0}$C the concentration of electronegative impurities was under
1.9$\times10^{-9}$ equiv. O$_{2}$ \cite{BAR92,BAR93}. The content
of electronegative impurities in argon was monitored with a
two-phase detector (1) \cite{BAR93,BARA93}. The absorber vessel,
made of high-temperature steel, was a cylinder 50 mm in diameter
and 300 mm high filled with the titanium sponge. The argon flow
rate during the purification phase was 0.6 m$^{3}$/h.

\subsection{Passive shielding}
The 15 cm thick passive shielding of lead was installed on a
concrete platform. Note that the measured radon activity in the
ambient air was 200 Bq/m$^{3}$ (for comparison, the radon activity
in the other halls of the Gran Sasso underground laboratory was 40
Bq/m$^{3}$). To suppress radon diffusion from the environment the
nitrogen vapor from the detector's cooling system was flushed
through a heat exchanger, heated to 20 $^{0}$C and directed inside
the passive shielding. With the same aim in mind hollow organic
glass blocks were placed next to the detector inside the passive
shielding (in order to reduce the volume of air inside the
shielding). As a result, the background due to $^{222}$Rn near the
detector (inside the passive shielding) was suppressed to a level
of $<$ 2 Bq/m$^{3}$ (before installation of the organic glass
blocks it was 50 Bq/m$^{3}$). The passive neutron shielding
included a 25 cm thick water layer to thermalize neutrons and a 1
cm thick layer of boric acid (H$_{3}BO_{3}$) powder to capture the
thermalized neutrons. Preliminary estimates showed that such
shielding should provide an almost tenfold suppression of the
background due to high-energy gamma's produced from the neutron
capture in the detector. The experimental test demonstrated that
the suppression factor for energies above 4 MeV was approximately
eight.

\subsection{Electronics and data readout system}

\begin{figure}
\begin{center}
\includegraphics[width=8cm]{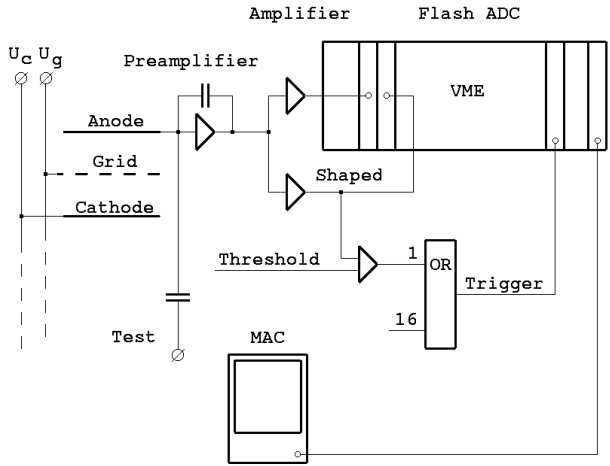}
\caption{Sketch of electronics and data-readout system.}
\label{ELECTRON}
\end{center}
\end{figure}

A diagram of the electronics and data readout system used in the
experiment is shown in Fig.~\ref{ELECTRON}. The outputs from all
15 anodes go to 15 ceramic feedthroughs. These signal
feedthroughs, mounted on the lid of the liquid argon chamber ((1)
in Fig.2), were connected to the charge sensitive preamplifiers
located on the upper lid of the detector. Each electronic channel
consisted of a charge sensitive preamplifier, an amplifier, and an
analog to digital converter (ADC) with a 20 MHz sampling
frequency. Since each channel recorded both shaped ($\tau _{sh}$=
3.6 $\mu $s) and unshaped (no filtering) signals, 30 flash ADCs
were used. Spatial information on the events was obtained by
analyzing the shape of the unfiltered pulses. A Macintosh computer
with the MACUA 1 data acquisition software was used for both
recording the data and controlling the electronics. The programs
for raw data histograming and an event display were written in the
same software environment. The trigger for data collection
required that at least one anode signal exceeded the 700 keV
threshold. Each trigger caused digitized signals from all anodes
to be written to a data tape. Such a universal triggering pattern
allowed the use of the power of pulse shape analysis to reduce the
background and allowed the investigation of single electron
spectra and the extraction of limits on $^{42}$Ar and $^{222}$Rn
content in liquid argon (see Section 3).

\subsection{Calibration procedure}

\begin{figure}
\begin{center}
\includegraphics[width=9cm]{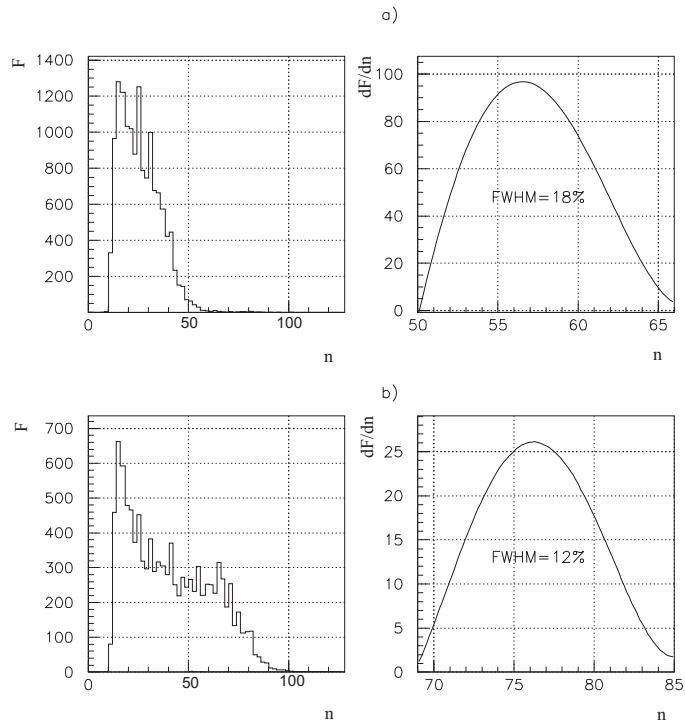}
\caption{Calibration spectra for (a) $^{22}$Na and (b) $^{88}$Y.
dF/dn is the first derivative of the polynomial fit to the
calibration spectrum; F(n) - counts number per one channel; \ n -
channel number.}
\label{Calibr}
\end{center}
\end{figure}

The detector filled with liquid argon was calibrated with external
gamma sources. The calibration was aimed at (a) providing a
relationship between the amount of energy deposited in the
detector's fiducial volume and the ionization signal amplitude
recorded by the readout electronics and (b) determining and
monitoring the stability and energy resolution of the detector.
The response of the electronics was regularly tested for linearity
and stability with a voltage pulse generator. The integral
nonlinearity was $<$ 1 \% throughout the dynamic range. A small 2
\% drift of the amplifiers gains was observed over the period of
data taking and was corrected using the pulse generator data. Two
$\gamma $-ray sources were selected for the calibration: $^{22}$Na
(T$_{1/2}$ = 2.6 yr, E$_{\gamma }$ = 1275 keV) and $^{88}$Y
(T$_{1/2}$ = 106.6 days, E$_{\gamma}$= 1836 keV). Since the
photoelectric effect probability depends heavily on the atomic
number of the element (P$_{ph}\sim Z^{5}$) and the atomic number
of argon is relatively small (Z = 18), it is virtually impossible
to use the photoelectric effect to calibrate the liquid argon
detector. However, the energy spectrum of Compton electrons is
appropriate for this purpose if the source intensity is high
enough to allow the fitting of the high energy edge of the Compton
spectrum followed by its differentiation. This approach was used
in the calibration procedure. A fifth degree polynomial function
was chosen to fit the spectrum. The peak of the first derivative
of the fit function corresponds to the maximum energy of the
Compton electrons. For $^{22}$Na and $^{88}$Y this energy is 1063
and 1610 keV, respectively. Each channel was calibrated
independently. Typical single channel calibration spectra of
$^{22}$Na and $^{88}$Y sources are shown in Fig.~\ref{Calibr}.
The full width at half-maximum (FWHM) of the distribution for the
channel shown was about 18 \% for 1063 keV ($^{22}$Na) and 12 \%
for 1610 keV ($^{88}$Y). The resolution of the other channels
varied from this value by no more than $\pm $10 \%. It is worth
noting that the relative energy resolution changed with energy as
1/E rather than $1/\sqrt{E}$. This is because the major
contribution to the energy resolution is capacitive noise in the
electronics readout chain (note the high value of the electrode's
capacitance - 500 pF) and the {"}microphone effect{"}. The latter
is caused by mechanical vibrations of the anode and grid wires in
the boiling liquid argon resulting in oscillation of the electrode
capacitance value. The extrapolation of the $^{22}$Na and $^{88}$Y
calibration results gives a 6 \% energy resolution (FWHM) at the
energy of the $^{100}$Mo $\beta \beta $ decay transition (3033
keV). The calibration procedure with the sources described above
was repeated every 300 h during the measurements. No deviations
from the initially measured values were observed over the course
of data taking.

\section{Physics results}
\begin{figure}
\begin{center}
\includegraphics[width=14cm]{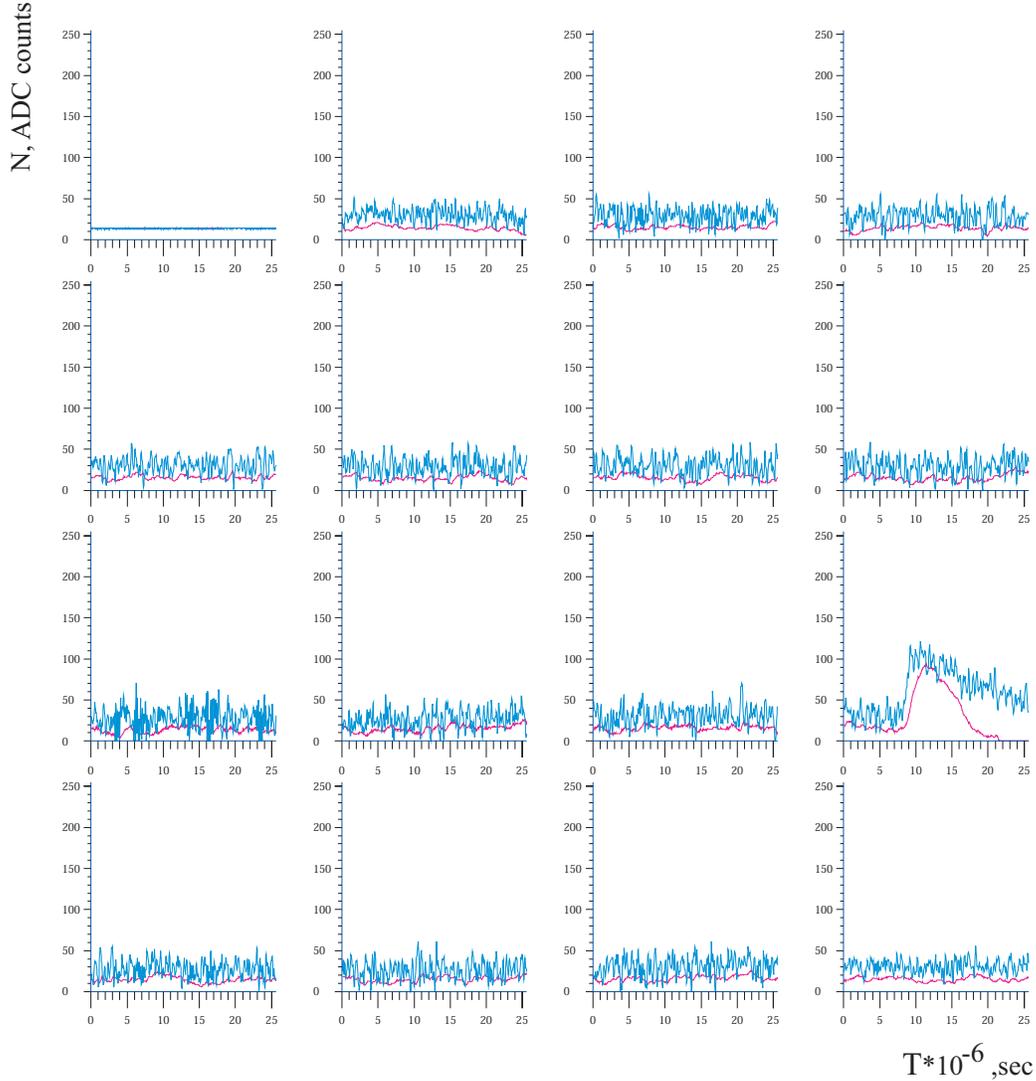}
\caption{A single-electron event (channel \# 12 is fired). The
signals are shown for all the 15 ADC channels: shaped (with RC
time constant T = 3.6 $ \mu $s, solid curve) and unshaped signals.
The channels are numbered from top left to bottom right. Channel
\# 1 was disconnected from the anode circuitry and used to monitor
the liquid nitrogen level and argon pressure in the sensitive
volume.}
\label{Single}
\end{center}
\end{figure}

\begin{figure}
\begin{center}
\includegraphics[width=14cm]{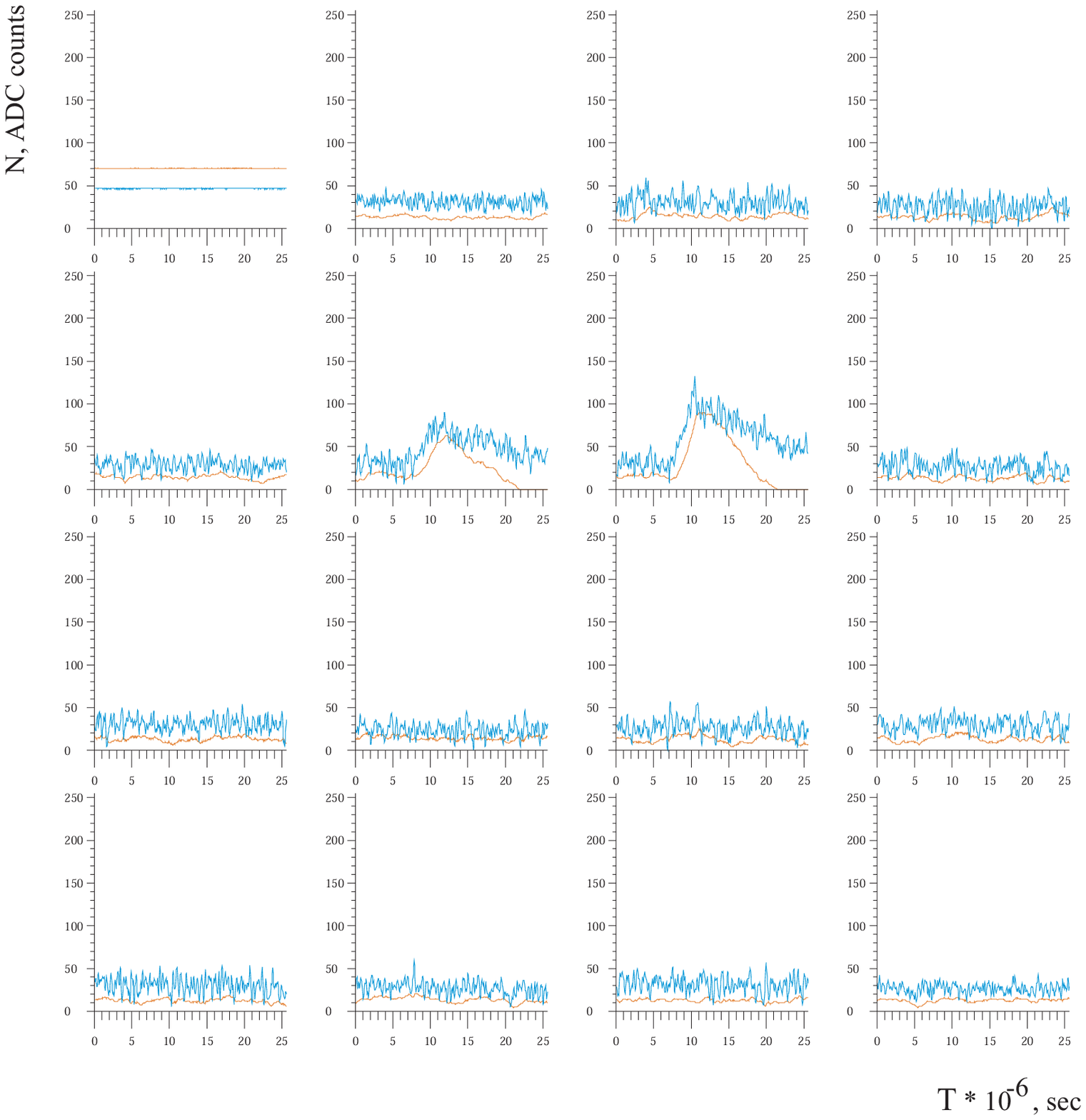}
\caption{A two-electron event (channels \# 5 and \# 6 are fired).
The signals are shown for all the 15 ADC channels: shaped (with RC
time constant T = 3.6 $ \mu $s, solid curve) and unshaped signals;
the channels are numbered from top left to bottom right. Channel
\# 1 was disconnected from the anode circuitry and used to monitor
the liquid nitrogen level and argon pressure in the sensitive
volume.}
\label{Double}
\end{center}
\end{figure}

A typical single electron event (only one anode signal above the
threshold) is shown in Fig.~\ref{Single}. Two-electron events
(events with two neighboring anode signals with a time difference
$<$ 0.6 $\mu$s) were selected as candidate events for the double
beta decay of $^{100}$Mo (Fig.~\ref{Double}).

\begin{figure}
\begin{center}
\includegraphics[width=10cm]{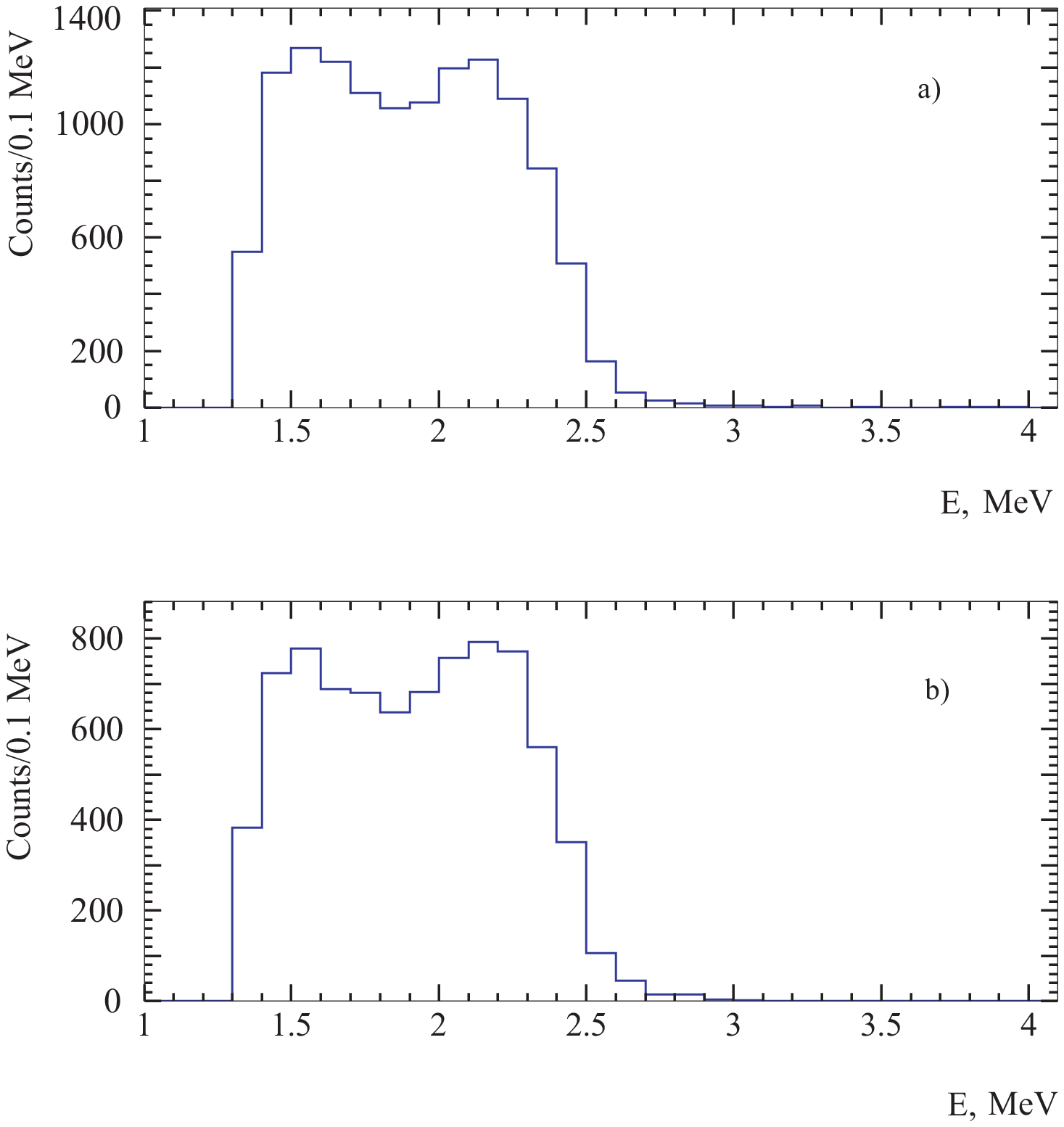}
\caption{Electron sum energy spectra of two-electron events. (a)
Enriched molybdenum (842.2 $ kg \times$ h) and (b) natural
molybdenum (592.9 $ kg\times$ h)}
\label{all-double}
\end{center}
\end{figure}

\begin{figure}
\begin{center}
\includegraphics[width=9cm]{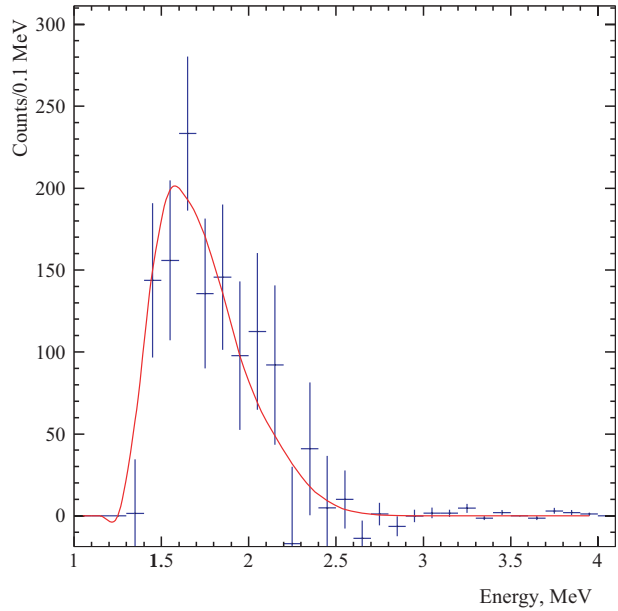}
\caption{Spectrum obtained as a result of subtraction of
natural-Mo two-electron spectrum from $ ^{100}Mo $ two-electron
spectrum. The curve shows theoretical spectrum of $^{100}$Mo
$\beta \beta (2\nu)$ events with $T_{1/2}=7.2\times 10^{18}$ yr.}
\label{2v}
\end{center}
\end{figure}

Figure ~\ref{all-double} shows the electron sum energy spectra of
these events for the sections with enriched and natural
molybdenum. The energy threshold is 0.8 MeV for the first electron
in the event and 0.5 MeV for the second. The detection efficiency
of each $\beta \beta $ decay mode was calculated by Monte Carlo
using the GEANT 3.21 program \cite{GEAN94}. The final analysis
involved the results from all experimental runs which corresponded
to 842.2 kg$\times $h of measurements for $^{100}$Mo and 592.9
kg$\times $h for natural molybdenum. The difference spectrum of
two-electron events (enriched -- natural) shown in Fig.~\ref{2v}
is due to the $\beta \beta(2{\nu})$ decay of $^{100}$Mo.
\bigskip
 \\ 0${\nu }$-mode
\bigskip
 \\To reduce the background the energy threshold for each electron of
a pair was set at 1 MeV. The energy range (2.8--3.1) MeV has been
studied with an additional selection on signal shape. As a result
6 events in the enriched molybdenum and 4 events in the natural Mo
(i.e. 5.8 events if recalculated for 848.2 kg$\times$h) have been
detected. Using the calculated detection efficiency (6.9\%) one
can obtain a limit on the $\beta\beta(0\nu)$ decay mode of
$^{100}$Mo,
\bigskip
 \\ $T_{1/2}>8.4(4.9)\times 10^{21}$ yr at 68 \% (90 \%) C.L.                         (1)
\bigskip
 \\ 0$\nu \chi^{0}$-decay
\bigskip
 \\The energy interval (2.3-3.0) MeV has been investigated. 1613
events for the enriched Mo foils and 1577 events for the natural
foils (rescaled for 848.2 kg$\times$h) have been obtained. For the
efficiency of 5.7\% we have obtained the limit,
\bigskip
 \\ $T_{1/2}>4.1(3.2)\times 10^{20}$ yr at 68 \% (90 \%) C.L.                          (2)
\bigskip
 \\ 2 $\nu $-decay
\bigskip
 \\Events have been analysed in the energy interval (1.4-2.4) MeV where the signal to
background ratio is maximal. After background subtraction 1140
$\pm $ 146 events remained. Using the calculated detection
efficiency of the $\beta\beta($2$\nu)$decay of $^{100}$Mo (2.2 \%)
one gets the half-life:
\bigskip
\\ $T_{1/2}=[7.2{\pm}0.9(stat){\pm}1.8(syst)]{\times}10^{18}$ yr                      (3)
\bigskip
\\The systematic error is mainly due to possible contributions of
radioactive impurities in the foils.

\begin{figure}
\begin{center}
\includegraphics[width=16cm]{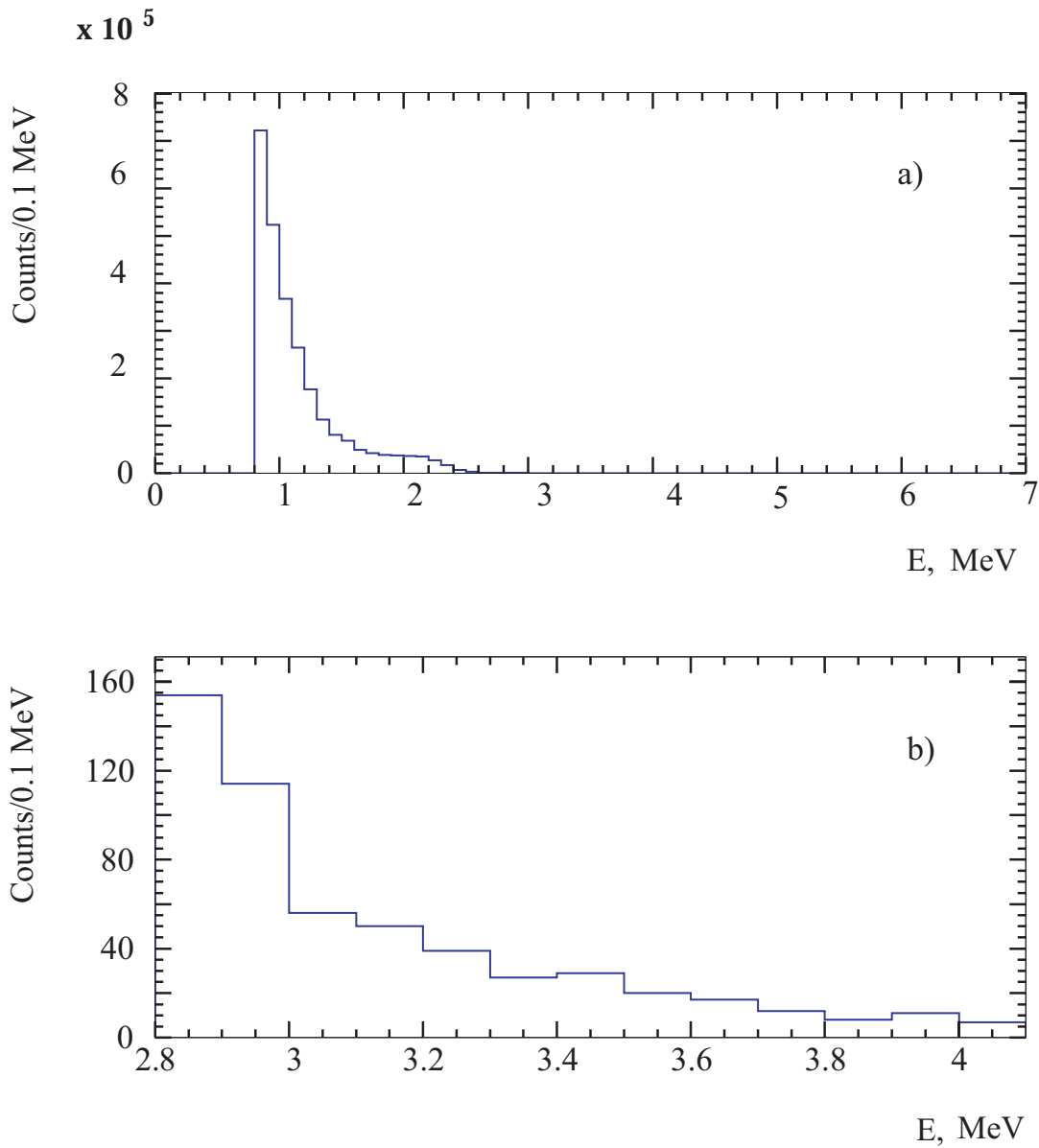}
\caption{Energy spectrum of single-electron events (all detector
sections summed). (a) The entire dynamic range of the detector,
(0--7) MeV; (b) zoomed view ((2.8--4.1) MeV energy interval).}
\label{all-single}
\end{center}
\end{figure}

The spectrum of single-electron events obtained over 2706 h of
data taking is shown in Fig. ~\ref{all-single}. It was used to
estimate the background in Mo-foil cathodes and to determine
$^{222}$Rn and $^{42}$Ar concentrations in liquid argon. These
radioactive isotopes constitute a potential danger for future
experiments utilizing large quantities of liquid argon (e.g.
ICARUS \cite{RUB01}). 145 and 201 events were observed in the
energy intervals (3.0--3.3) MeV (the end point of the $^{214}$Bi
beta spectrum) and (3.0--3.5) MeV (the end point of the $^{42}$Ar
beta spectrum) respectively. The contribution to the background in
these regions from the high energy plateau was estimated from the
high energy part of the spectrum above 3.5 MeV . The maximum
number of counts due to $^{222}$Rn and $^{42}$Ar in the mentioned
energy regions were found to be 70 and 127 events respectively.
The detection efficiency for these events was calculated by Monte
Carlo assuming that the decay products were either uniformly
distributed in liquid argon or were deposited on the cathodes. The
difference in the calculated efficiency between the two cases was
less than 10 \%. The efficiencies were 0.025 \% for $^{222}$Rn and
1 \% for $^{42}$Ar. The corresponding upper limit on the
$^{222}$Rn activity in liquid Ar is 1.2$\times $10$^{-3}$ Bq/kg at
90 \% C.L. The upper limit on $^{42}$Ar content is 4.3$\times
$10$^{-21}$ g/g at 90 \% C.L..

\section{Conclusions}

The liquid argon ionization detector was installed, commissioned
and demonstrated good performance characteristics. The high level
of purity of liquid argon (1.9$\times 10^{-9}$ equiv.O$_{2}$)
needed for a successful operation of the detector was achieved.
The data readout system developed made it possible to select
reliably double-beta decay events and separate them from numerous
background events. The two experimental runs carried out in 2000
showed that the passive neutron shielding had suppressed the high
energy background in the fiducial volume by a factor of 8. The
results obtained over a period from 1996 to 2000 are the
following:
\\1) The $\beta \beta (2\nu)$ decay of $^{100}$Mo
with a half-life of [7.2$\pm 0.9(stat)\pm 1.8(syst)]\times
10^{18}$ yr was observed \cite{ASH01}.
\\2) The lower limits on the half-lives for 0$\nu $ and 0$\nu \chi^{0}$
modes of the $^{100}$Mo ${\beta \beta }$ decay were put at
8.4(4.9)$\times 10^{21}$ and 4.1(3.2)$\times 10^{20}$ yr
respectively at 68 \% (90 \%) C.L.\cite{ASH01}.
\\3) A new experimental limit of $< 1.2\times 10^{-3} $ Bq/kg (90 \% C.L.)
 was obtained for the $^{222}$Rn activity in liquid argon.
\\4) The content of radioactive $^{42}$Ar in the Earth's atmosphere was found
to be less than 4.3 $\times 10^{-21}$ g/g (90 \% C.L.). \\ It
should be noted that the two limits are the world's best results
for the content of $^{222}$Rn and $^{42}$Ar in liquid argon which
should help understand potential backgrounds in future detectors
using large quantities of liquid argon as detection media.

\section*{Acknowledgement}
We are grateful to M.Balata (LNGS) for his help in maintaining
the liquid nitrogen cooling system of the experiment and for
fruitful discussions. We are also thankful to the LNGS workshop
staff, in particular to L.Marelli, B.Romualdi, A.Rotilio and
E.Tatananni for regular and timely technical support.


\begin{thebibliography}{References}
\bibitem{KLAP98}
H.V. Klapdor-Kleingrothaus, J. Hellmig, M. Hirsch, J. Phys. G 24
(1998) 483.
\bibitem{FEAS01}
A. Faessler and F. Simkovich, Prog. Part. Nucl. Phys., 46 (2001)
233.
\bibitem{VER02}
J.D. Vergados, Phys. Rep., 361 (2002) 1.
\bibitem{KAJ01}
T. Kajita and Y. Totsuka, Rev. Mod. Phys. 73 (2001) 85.
\bibitem{CLEV98}
B.T. Cleveland et al., Astrophys. J. 496 (1998) 505.
\bibitem{ABD99}
J.N. Abdurashitov et al., Phys. Rev. C60 (1999) 055801.
\bibitem{HAM99}
W. Hampel et al., Phys. Lett. B447 (1999) 127.
\bibitem{AHM01}
Q.R. Ahmad et al., Phys. Rev. Lett. 87 (2001) 071301; ibid 89
(2002) 001301; ibid 89 (2002) 001302.
\bibitem{FUK01}
S. Fukuda et al., Phys.Rev. Lett.86 (2001) 5651; ibid 86 (2001)
5656.
\bibitem{EGU03}
K. Eguchi et al., Phys. Rev. Lett. 90 (2003) 041801.
\bibitem{BAR86}
A.S. Barabash, A.I. Bolozdynya and V.N. Stekhanov, Preprint of
Inst. of Experimental and Theoretical Physics, Moscow, 1986, \#
154.
\bibitem{ASH01}
V.D. Ashitkov, A.S. Barabash, S.G. Belogurov, et al., JETP Lett.
74 (2001) 529.
\bibitem{BAR92}
A.S. Barabash and V.N. Stekhanov, Nucl. Instrum. Methods Phys.
Res., Sect. A, 316 (1992) 51.
\bibitem{BAR93}
A.S. Barabash and V.N. Stekhanov, Nucl. Instrum. Methods Phys.
Res., Sect. A, 327 (1993) 168.
\bibitem{BARA93}
A.S. Barabash and A.I. Bolozdynya, Zhidkostnye ionizatsionnye
detektory (Liquid Ionization Detectors), Moscow: Energoatomizdat,
1993.
\bibitem{GEAN94}
GEANT 3.21. Detector Description and Simulation Tool, CERNLIB,
1994.
\bibitem{RUB01}
A. Rubbia, Nucl. Phys. B (Proc. Suppl.) 91 (2001) 223.

\end{thebibliography}
\end{document}